\documentclass[prl,twocolumn,lengthcheck,superscriptaddress,showpacs,preprintnumbers,amsmath,amssymb]{revtex4}
\usepackage{graphicx}% Include figure files
\usepackage{dcolumn}% Align table columns on decimal point
\usepackage{bm}% bold math
\newcommand{\etal}{\emph{et al. }}
\begin{document}
%\preprint{\tt Not for circulation}

\title{Mean-field phase diagram of cold lattice bosons in disordered potentials}

\author{P. Buonsante}
\affiliation{Dipartimento di Fisica, Politecnico di Torino Corso Duca degli Abruzzi 24, I-10129 Torino (ITALIA)}%
\affiliation{Jack Dodd Centre for Photonics and  Ultra-Cold Atoms, Department of Physics, University of Otago, P.O. Box 56, Dunedin, New Zealand.}
 \author{V. Penna}
 \affiliation{Dipartimento di Fisica, Politecnico di Torino Corso Duca degli Abruzzi 24, I-10129 Torino (ITALIA)}%
 \author{A. Vezzani}
 \affiliation{Dipartimento di Fisica, Universit\`a degli Studi di Parma and C.N.R.-I.N.F.M., Parco Area delle Scienze 7/a, I-43100 Parma (ITALIA)}
\author{P.B. Blakie}
\affiliation{Jack Dodd Centre for Photonics and  Ultra-Cold Atoms, Department of Physics, University of Otago, P.O. Box 56, Dunedin, New Zealand.}

\date{\today}

\begin{abstract}
We address the phase diagram of the disordered Bose-Hubbard model 
that has been  realized in several recent experiments in
terms of optically trapped ultracold bosons. We show that a 
a systematic description of all of the expected quantum phases can
be obtained at both zero and finite temperature via a  
site-dependent decoupling mean-field approach. 
Also, we relate the boundaries of 
the Mott insulating phase to an off-diagonal non-interacting 
Anderson model whose spectral features
provide a new avenue for determining the debated nature of the 
phase surrounding the Mott lobes.
Our approach is simple yet effective, and scalable to
systems with experimentally relevant sizes and features.
\end{abstract}

\maketitle

In 1998 Jaksch \etal\cite{A:Jaksch} demonstrated that the Bose-Hubbard (BH) model \cite{A:Fisher} could be accurately realized with a degenerate Bose gas in an optical lattice. Testament to the rapid pace of experimental development in this field, the defining superfluid-insulator quantum phase transition of this model was first observed by Greiner \etal \cite{A:Greiner02} in 2002. 
The control over atoms afforded by optical lattices has served to initiate a broad range of investigations with bosonic and fermionic atoms \cite{pack1,A:Ospelkaus}. 

While optical lattices naturally provide a defect-free periodic potential that allows precise control over the relative strengths of interactions and tunneling, there is much interest relating to the interplay between interactions and disorder in the BH model. Indeed, experiments are currently developing in this direction and have demonstrated several approaches for engineering disorder in optical lattices, such as: using laser speckle fields \cite{pack2}; an additional incommensurate lattice \cite{A:Roth03,A:Damski2003,A:Fallani}; or using a distinguishable atom species to act as a randomly distributed set of impurities \cite{pack3,A:Ospelkaus}.

As it was first demonstrated in Ref. \cite{A:Jaksch}, degenerate bosonic atoms in an optical lattice are described by the Bose-Hubbard (BH) Hamiltonian, 
%%%change: introduced n_m in eq (2), so that it is evident that H_m 
%%% is a function of the number, and its eigenstates are number states.
%%% Hence I include a definition of n_m in the sentence following eq. (2)
\begin{eqnarray}
\label{E:BHH}
H & = & \sum_{m=1}^M H_m - J \sum_{m,m'} a_m^\dag A_{m\,m'} a_{m'}, \\
H_m & = & \frac{U}{2} n_m (n_m-1) + (v_m -\mu) n_m, 
%a_m^\dag a_m^\dag a_m a_m + (v_m -\mu) a_m^\dag a_m, 
\end{eqnarray}
originally introduced as a toy model of superfluid $^4$He in porous media \cite{A:Fisher}.  The operators $n_m = a_m^\dag a_m$, $a_m$ and $a_m^\dag$  respectively count, destroy and create  bosons at lattice site $m$, and obey canonical commutation rules $[a_m,a_{m'}^\dag]= \delta_{m \, m'}$. The chemical potential $\mu$  is related to the total boson population $N=\sum_m n_m$. 
The Hamiltonian parameters, namely the boson-boson (repulsive) interaction $U>0$, and the hopping amplitude across neighbouring sites $J$  are directly related to  the atomic scattering length, and lattice depth \cite{A:Jaksch}. The {\it adjacency} matrix $A_{m\,m'}$ appearing in the hopping term allows a simple algebraic description of the lattice structure, being finite if  sites $m$ and $m'$ are nearest neighbours and zero otherwise. 
The local potential $v_m$ at site $m$ is related to the features of the effective potential \cite{A:Jaksch}. Here this quantity will be random to  realize disorder.

As discussed in the seminal paper by Fisher \etal\cite{A:Fisher}, the presence of disorder enriches the phase diagram of the BH model --- that in the {\it pure } case ($v_m=0$) consists of an extended superfluid region and a series of Mott-insulator lobes ---  with a further {\it Bose-glass} phase. 
Similar to the  Mott insulator, the Bose glass phase is characterized by the absence of superfluidity, however has a finite compressibility (or gapless spectrum) like the superfluid phase. 
A representative list of techniques used to
investigate the disordered BH  phase diagram
includes field-theoretic approaches
\cite{A:Fisher,pack4,A:Svistunov1996,A:Pazmandi},
decoupling (or Gutzwiller) mean-field
approximations
\cite{A:Sheshadri,A:Krauth1992,A:Sheshadri1995,A:Damski2003,A:Krutitsky},
quantum Monte Carlo  simulations
\cite{pack5,A:Lee2004}
and others
\cite{pack6,A:Freericks1996,A:Pai,A:Rapsch,pack7}.
Nevertheless, several aspects of the problem are still subject of active debate, such as a precise characterization of the different phases \cite{pack7}, the issue of the direct transition from MI to SF phase \cite{A:Fisher,A:Pai,A:Svistunov1996,A:Pazmandi,A:Lee2004}, and the phase diagram at finite temperature \cite{A:Krutitsky}.

In this paper we show that a site-dependent 
%decoupling 
%%%change: It seemed to me that ``decoupling'' was superfluous and eroding precious real estate. 
mean-field approach \cite{A:Sheshadri1995,A:LobiMF} captures all of the essential features of the phase diagram of the disordered BH model, 
%%%change: Vittorio feels that ``at both X and Y'' is more correct/usual than ``bot at X and Y''. Is it?
at both  zero and finite temperature. As summarized in the legend of figure \ref{F:fT}, the different phases, namely the Mott insulator (MI), the Bose glass (BG), the superfluid (SF) and --- at finite temperature --- the normal fluid (NF) are characterized by the value of three quantities, i.e. the superfluid fraction $f_{\rm s}$, the compressibility $\kappa$ and the condensate fraction $f_{\rm c}$. For  simplicity we present results for a translationally invariant 1D
%one-dimensional 
%%%change: again here I am being greedy with ``real estate''. I'm pretty sure that 1D is an accepted abbreviation. I've seen it even in titles. I am going to use it throughout the manuscript, if you agree. At worst we can define it once and for all here but, as I say, I think it is not needed.
lattice with random on-site potentials $v_m$ uniformly distributed in $[-\Delta, \Delta]$. 
However we emphasize that the mean-field approach lends itself to more general situations, such as higher dimensional systems, different realizations of disorder and realistic trapping potentials \cite{N:elsewhere}.
%We note however that a more detailed analysis considering higher dimensional systems, different realizations of disorder and realistic trapping potentials is within the capabilities of the mean-field approach \cite{N:elsewhere}. \st
% A more detailed analysis considering higher dimensional systems, different realizations of disorder and realistic trapping potentials will be reported elsewhere \cite{N:elsewhere}.
%%%change: I tried to give a more positive spin to this sentence, as suggested by Alessandro. He was worried about a referee saying ``if you can do this, include it in this Letter''. I tried to convey the fact that this can be done, yet it is still underway. I do not know whether ``within the capabilities of'' is really correct.
 Before describing our results we redefine the parameters using $U$ as the energy scale,  so that henceforth $J$, $\mu$, $v_m$, $\Delta$ are to be intended as $J/U$, $\mu/U$, $v_m/U$, $\Delta/U$.  

The superfluid fraction is determined by the stiffness of the system under phase variations, 
\begin{equation}
\label{E:fs}
f_{\rm s} = \lim_{\theta \to 0} \frac{E_\theta -E_0}{\langle N \rangle J \theta^2}
\end{equation}
%where $N$ and $E_\theta$ are the total number of bosons and the energy of the system with twisted boundary conditions, respectively.  The latter is obtained by introducing the so called {\it Peierls phase factors} in the kinetic term of Hamiltonian (\ref{E:BHH}).
where $\langle \cdot\rangle$ denotes thermal average and $E_\theta$ is the energy of the system with twisted boundary conditions.  The latter are obtained by introducing the so called {\it Peierls phase factors} in the kinetic term of Hamiltonian (\ref{E:BHH}). In 1D  this amounts to setting $A_{m\,m'} = e^{-i \theta} \delta_{m\,m'+1}+e^{i \theta} \delta_{m\,m'-1}$ \cite{A:Roth03}. The compressibility is defined as $\kappa = \partial_\mu N = \beta (\langle N^2\rangle-\langle N \rangle^2)$, where $\beta = U/k_BT$ is the inverse temperature. Finally, the condensate fraction, $0\leq f_{\rm c}\leq 1$ is defined as the largest eigenvalue of the one body density matrix $\rho_{m\,m'}=\langle a_m^\dag a_{m'} \rangle/N$ \cite{A:Roth03,A:Penrose}. 
%%%change: lots of changes here. I moved some definitions around, fixed some problems with the definition of \kappa and used 1D in place of one-dimensional.

The decoupling mean-field approach \cite{A:Sheshadri,A:Sheshadri1995} results from the approximation $a_m^\dag a_{m'} \approx a_m^\dag \alpha_{m '}+ \alpha_m^* a_{m'} - \alpha_m^* \alpha_{m'}$, with $\alpha_m = \langle a_m\rangle$.  This turns Hamiltonian (\ref{E:BHH}) into  ${\cal H}=\sum_{m=1}^M {\cal H}_m + h$, where ${\cal H}_m  =  H_m - J (\gamma_m a_m^\dag  + \gamma_m^* a_m)$ and
\begin{equation}
\label{E:gammam}
\gamma_m= \sum_{m'=1}^M A_{m\,m'} \alpha_{m'}, \;\; h \!=\! J\!\sum_{m=1,m'}^M \alpha_m^* A_{m\,m'} \alpha_{m'}
\end{equation}
Since the  mean-field Hamiltonian ${\cal H}$ is the sum of on-site terms ${\cal H}_m$, the original problem is reduced to a set of  problems involving quantities relevant to neighbouring sites \cite{A:LobiMF,A:Bru}
\begin{equation}
\label{E:map}
\alpha_m = \frac{{\rm tr}(a_m e^{-\beta {\cal H}_m})}{{\rm tr} \,e^{-\beta {\cal H}_m}}.
\end{equation}
Indeed, according to Eq.~(\ref{E:gammam}), the local Hamiltonian ${\cal H}_m$ depends on the $\alpha_{m'}$'s at sites $m'$ adjacent to $m$, thus through the $\{\alpha_m\}$
 spatial correlations in the system are approximately included in the decoupling approach.  One useful measure of these spatial correlations is the  one-body density matrix,   $\rho_{m\, m'} = [(\langle n_m\rangle - |\alpha_m|^2)\delta_{m\, m'} + \alpha_m^* \alpha_{m'}]/N$, as defined above.
 Note that the set of $\alpha_m$'s characterizing the state of the system can be seen as a stable fixed point of the map defined by Eq.~(\ref{E:map}). An easily found fixed point corresponds to the choice $\alpha_m =0$ for all sites. In Ref.~\cite{A:LobiMF} the stability of such fixed point is studied for $T>0$. In the  $T=0$ case Eq. (\ref{E:map}) turns into $\alpha_m = \langle\Psi|a_m|\Psi \rangle = \langle\psi_m|a_m|\psi_m \rangle$, where $|\Psi\rangle = \prod_{m=1}^M |\psi_m \rangle$ and  $|\psi_m \rangle$ are the ground-states of the entire system and of ${\cal H}_m$, respectively. Note that in this limit the fixed point $\alpha_m =0$ corresponds to the number-squeezed ground state typical of the MI phase. Indeed, since ${\cal H}_m = H_m$ one easily gets $|\psi_m \rangle = |\nu_m \rangle$, where $n_m |\nu_m \rangle = \nu_m |\nu_m \rangle$ and $\nu_m = \lceil \mu-v_m\rceil \in {\mathbb N}$. Making use of first order perturbation theory it is possible to show that this MI phase is stable for $J < J_{\rm c} = |q_M|^{-1}$, where $q_M$ is the maximal eigenvalue of the matrix of entries $Q_{m\, m'} = g_m(\mu) A_{m\, m'}$, with $g_m(\mu)=g(\mu-v_m)$ and $g(x) = (x+1)/[(\lceil x\rceil-x)(x-\lfloor x\rfloor)]$ \cite{N:elsewhere}. 
Note that for the $\alpha_m=0$ fixed point $f_{\rm s} = \kappa =  0$, while $f_{\rm c} \propto M^{-1} \to 0$, as expected for a MI state (see legend, Fig. \ref{F:fT}). As soon as $\alpha_m \neq 0$ the local ground state is not a number state any more, $|\psi_m \rangle = \sum_{\nu = 0}^\infty c_{m \nu} |\nu \rangle$, and the system enters a compressible phase. Also, it can be shown that $f_{\rm c} \propto \overline{|\alpha_m|^2}$, where the bar denotes average over the lattice sites. Hence the compressible phase found for $J\geq J_{\rm c}$ has a finite condensate fraction, i.e. long range correlations. Generally the superfluid fraction $f_{\rm s}$  has to be evaluated  numerically  \cite{N:elsewhere}. 
In the following we show that on 1D systems the boundary of the SF region is simply  related to the vanishing of $\alpha_m$ at some site of the lattice.
%%%change: fixed this thing about the sf fraction.
Before discussing our results, we rapidly review the {\it pure} BH model. Since $v_m = 0$, $Q = g(\mu) A$, and  the known equation for the boundary of the Mott lobes, $J_{\rm c} = [2  g(\mu)]^{-1}$ \cite{A:Fisher}, is easily recovered. As we mention above, $\kappa>0$ as soon as $\alpha_m>0$. Furthermore, it is not hard to show that $f_{\rm c} = f_{\rm s} = |\alpha_m|^2/\langle n_m \rangle$, where the first equality is true in the thermodynamic limit $M\to \infty$.

Let us now consider $v_m$ uniformly distributed in $[-\Delta, \Delta]$, focusing first of all on the boundaries of the MI phases. We begin by noting that  the matrix $Q$ whose maximal eigenvalue gives the critical value $J_c$ can be related to the Hamiltonian for an off-diagonal Anderson model whose random hopping amplitudes have an unusual distribution \cite{N:odA} .
 This can be seen observing that $Q$ has the same spectrum as the symmetric matrix of elements $R_{m\,m'} = \tau_{m\,m'}  A_{m\,m'} = \sqrt{g_m(\mu)g_{m'}(\mu)} A_{m\,m'}$, describing noninteracting particles hopping across the sites of the lattice described by $A$ with random amplitudes given by $\tau$.
 The spectrum of $R$ can be analyzed for very large ($M\sim 10^6$) 1D systems using e.g. transfer matrix methods \cite{A:Politi}. It is easy to see that the evaluation of $J_{\rm c} = |q_M|^{-1}$ makes sense only for $\mu \in {\cal J_\nu}=(\nu+\Delta,\nu+1-\Delta)$, where $\nu$ is a non negative integer and $\Delta<1/2$. Indeed if $\mu \in {\cal I_\nu}=[\nu-\Delta,\nu+\Delta]$ there is the possibility (certainty for $M\to \infty$) 
%%%change: discarded ``which becomes'' from the parenthetic remark. It still works, isn't it?
that $\mu-v_m=\nu\in {\mathbb N}$ for one or more of the $v_m$'s. This means that $g_m(\mu)$ diverges, and $J_c = 0$. Hence, as expected, the Mott lobes are found only within the $\mu$ intervals ${\cal J_\nu}$ \cite{A:Fisher}, where the entries of $Q$ are always finite. 
We also note that if $g_m(\mu)$ is replaced by its average $\overline g(\mu)$, which discards the spatial correlations inherent in the Anderson model described by $R$, our approach reproduces the boundaries given in Refs. \cite{A:Fisher,A:Freericks1996,A:Krutitsky}.

\begin{figure}[t!]
\begin{center}
\includegraphics[width=8.5cm]{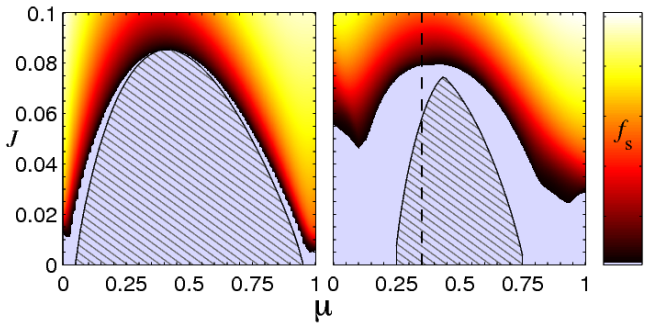}
\caption{\label{F:pd}  Phase diagrams for a disordered lattice comprising $M=100$ sites for two values of the strength of the random potential,  $\Delta = 0.05$ (left) and $\Delta = 0.25$ (right).  }
\end{center}
\end{figure}

\begin{figure}[t!]
\begin{center}
\includegraphics[width=8.5cm]{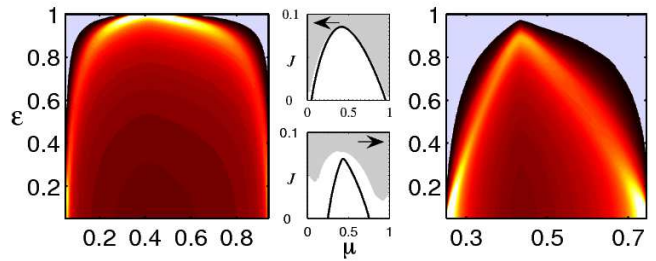}
\caption{\label{F:ids} Spectral density
  $\sigma(\epsilon)$ of $Q$ vs. $\mu$ for a lattice comprising $M=10^6$ sites. The leftmost (rightmost) panel corresponds $\Delta=0.05$ ($\Delta=0.25$).  The spectra are normalized so that $q_M$ corresponds to $\epsilon=1$. The small central panels show the corresponding phase diagrams, as indicated by the arrows. The gray areas are the SF regions, the solid black lines are the MI lobes.}
\end{center}
\end{figure}

\begin{figure}[t!]
\begin{center}
\includegraphics[width=8.5cm]{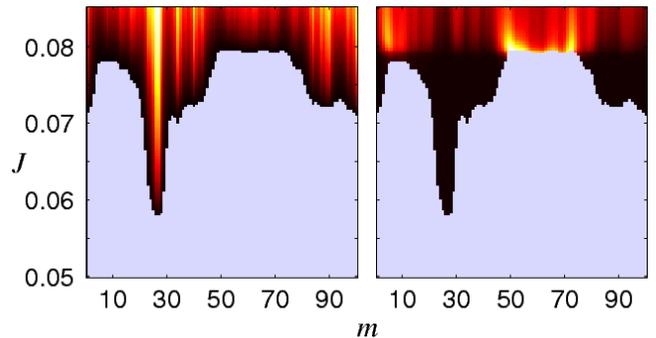}
\caption{\label{F:lattice} Square modulus (left) and phase differences (right) of $\alpha_m^{(\theta)}$  along the dashed line in the right panel of Fig~\ref{F:pd} ($\Delta = 0.25$, $\mu = 0.35$, $\theta = 10^{-3}$). }
\end{center}
\end{figure}

\begin{figure}[t!]
\begin{center}
\includegraphics[width=8.5cm]{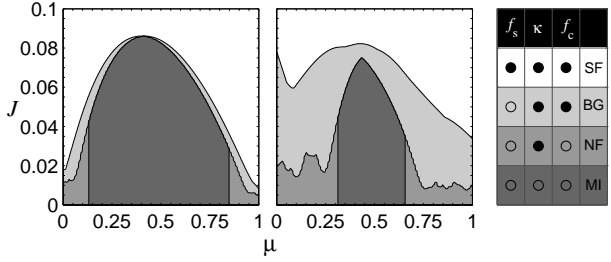}
\caption{\label{F:fT} The same phase diagrams as in Fig \ref{F:pd}. at a finite temperature $T=0.01 U/k_{\rm B}$. Filled and empty circles in the legend denote finite and zero values for the relevant quantity. }
\end{center}
\end{figure}

In what follows we discuss some numerical results obtained on 1D lattices  of size $M=100$ where periodic boundary conditions are assumed. We consider a fixed realization for  the ``profile'' of the disordered potential, $u_m\in [-1,\, 1]$ and obtain the actual potential at a given value of the strength $\Delta$  as  $v_m = \Delta u_m$. By averaging over disorder and considering large lattices we have verified that finite size effects are negligible in these results.
%%%change: here I simply removed a redundant sentence 
The values of $f_{\rm s}$ have been obtained evaluating $E_\theta$ with a self-consistent minimization procedure. We note that the same results can be derived
 from the self-consistent solution at $\theta =0$ after a perturbative expansion in $\theta$ \cite{N:elsewhere}. 
%%%change: I mentioned that one does not need to make two self-consistent calculations to determine f_s. This can be useful, above all on large systems.
 Fig. \ref{F:pd} shows the $T=0$ phase diagram  of the system in the region of the first Mott lobe for $\Delta = 0.05$ and $\Delta = 0.25$. Note that the extended uniformly-colored region in the lower part of both panels refers to vanishing $f_{\rm s}$, according to the colorbar.
%The colorplot refers to the values of $f_{\rm s}$. The darker the hue, the smaller the superfluid fraction. The  extended uniformly-colored region in the lower part of each panel corresponds to vanishing $f_{\rm s}$ (a colorbar is shown in Fig. \ref{F:lattice}). 
%%%change: I introduced a colorbar in Fig. 1, which greatly simplifies the description of the phase diagram.
The hatched regions correspond to the Mott lobes as evaluated computing the largest eigenvalue of the matrix $Q$. As expected, both $\kappa$ and $f_{\rm c}$ obtained from the numerical minimization of the mean-field energy vanish inside and are finite outside the hatched regions.  
These results clearly show that the disordered potential induces the appearance of a phase that is absent in the {\it pure} model, characterized by finite compressibility (and condensate fraction) but vanishing superfluidity. This phase is hence naturally identified with the Bose-glass (BG). 
In more detail, increasing the strength of the disorder $\Delta$ causes the BG to extend at the expense of the SF and  MI phases.
In Ref. \cite{A:Fisher} Fisher \emph{et al.} %conjecture 
%%%change: it seemed to me that either ``conjecture'' or ``suggest'' was superfluous here. I chose the latter because the former is used below (but avoiding repeated terms is probably an Italian mania). 
 suggest that the MI to SF transition always occurs through a BG phase, a conjecture that has been actively debated, e.g. see \cite{A:Pai,A:Svistunov1996,A:Pazmandi,A:Lee2004}. The results presented here suggest that, in the mean-field picture,  the MI and SF phases appear to be still connected near the tip of the Mott lobe for weak disorder (left panel of Fig. \ref{F:pd}). 
%It is worth recalling that, on 1D systems like the one we are considering, the present mean-field approach is not reliable in this region of the phase diagram, and hence the direct transition could be an artifact of the approximation \cite{A:Pazmandi,A:Rapsch}.  
Furthermore, our approach provides a new avenue for understanding the  presence of a BG phase separating the MI and the SF 
%on higher dimensional systems 
via the spectral features of the Anderson model associated to the matrix $Q$. 
%%%change: I changed ``described by'' into ``associated to'', and added ``on higher dimensional systems''. Also I changed the quantity plotted in fig.2, which now is the (non-integrated) spectral density of Q. A series of changes follow
This can be seen in Fig. \ref{F:ids}, showing the spectral  density $\sigma$ of $Q$ (or, equivalently, of $R$). Note indeed that for large disorder (right panel) $\sigma$ is always vanishing in the proximity of the band edge, whereas for small disorder (left panel) one can recognize a clearly different behaviour. In the vicinity of the tip of the lobe, where the SF and MI phases seem to be directly connected,  $\sigma$ has an evident peak, similar to what happens on a homogeneous 1D lattice.
We further observe that for 1D systems $f_{\rm s}$ vanishes as soon as the mean-field parameters vanish at two adjacent sites, e.g. $\alpha_{\overline m-1} = \alpha_{\overline m}=0$. Indeed in this situation one can verify that $E_\theta = E_0$ since $\alpha_m^{(\theta)} = |\alpha_m^{(0)}| e^{i \vartheta_m}$, where the parenthetic superscript refers to the value of the Peierls phase and $\varphi_m = \vartheta_m -\vartheta_{m+1} = \theta \,{\rm mod}(\overline m -m, M)$.  
  This can be derived from the self-consistency constraint $\gamma_m^{(\theta)}= e^{-i \theta} \alpha_{m-1}^{(\theta)}+e^{i \theta} \alpha_{m+1}^{(\theta)} $ observing that $\alpha_m^{(\theta)}$ inherits the phase factor from $\gamma_m^{(\theta)}=|\gamma_m^{(\theta)}| e^{i \vartheta_m} $   due to the specific form of  ${\cal H}_m$. Fig. \ref{F:lattice} clearly illustrates this phenomenon displaying what happens on the lattice while crossing the dashed line in the right panel of Fig. \ref{F:pd}, for $\theta = 10^{-3}$. As soon as $|\alpha_m^{(\theta)}|$ (left panel) vanishes at two or more adjacent sites, $f_{\rm s} = 0$ and  $\varphi_m$ (right panel) stops fluctuating and equals $\theta$ wherever it is defined.
Of course on higher dimensional systems the above argument does not apply, and one expects  the onset of superfluidity to be related to the percolation of the $\alpha_m$'s through the lattice \cite{A:Sheshadri1995}.

The last results we present here are the same phase diagram as in Fig. \ref{F:pd}, but at a finite temperature $T=0.01U/k_B$. As it can be seen in Fig. \ref{F:fT}, the values of $f_{\rm s}$, $\kappa$ and  $f_{\rm c}$ allow the characterization of four different phases. Strictly speaking, at finite temperatures the MI is replaced by a normal fluid (NF) which is always compressible. However at small temperatures some regions of the NF phase feature a compressibility so small that they can be considered (quasi) MI \cite{A:Sheshadri,A:LobiMF,A:Bru}. 
In particular our results show that at finite $T$ the BG cannot be characterized simply as a compressible non-superfluid phase. Indeed this is true also of the NF phase, where $f_{\rm s}=0$ because $\alpha_m = 0$ everywhere due to thermal fluctuations, even in the absence of disorder. The condensate fraction distinguishes between the NF and BG, i.e. the region $f_{\rm s}=0$  is comprised of BG phase with $f_{\rm c}\neq 0$ --- where the superfluidity is destroyed (mainly) by the presence of disorder --- and the NF phase with $f_{\rm c}= 0$ -- where the superfluidity is destroyed (mainly) by thermal fluctuations. As $T$ is increased the  NF dominates over the BG phase, while the (quasi) incompressible MI lobes shrink, until eventually only SF and NF phases remain.  We mention that the latter phase has glassy features according to the treatment described in Ref. \cite{A:Krutitsky}.

In summary, in this work we show that the site-dependent mean-field decoupling approach captures the phases of the disordered Bose-Hubbard model at both zero and finite temperature, and that in general three indicators should be taken into account to characterize the phase diagram. We observe that the boundaries of the MI lobes can be related to a non-interacting off-diagonal Anderson model. In particular, we present the phase diagrams of ideal one-dimensional systems for different values of the strength of the disorder, and suggest
that the specific nature of the transition from the MI to the SF phase might be related to the spectral features of the relevant Anderson model, which we investigate for lattice sizes up to $M=10^6$. Also, we observe that on 1D systems the superfluidity is destroyed as soon as the local mean-field parameters vanish at two adjacent sites, and sketch an argument explaining this.
We conclude observing that the site-dependent decoupling mean-field approach appears to be a very promising tool for investigating more general situations, including  different realizations of the disorder, higher dimensions and realistic features, such as the harmonic confinement that characterizes actual experiments \cite{pack2,pack8}.

{\bf Acknowledgments}. P.B. acknowledges a grant from the {\it Lagrange Project} - CRT Foundation and is grateful to the Jack Dodd Centre for the warm hospitality, as well as to P. Jain and F. Ginelli for useful suggestions.

%\bibliography{../../BIBTEX/biblio,./bose_glass,./BGL_notes,./bandstructbib}
%\bibliographystyle{../../BIBTEX/etal2}
\end{document}